\documentclass{DISproc}

\def\cascade{{\sc Cascade}}
\def\pythia{{\sc Pythia}}
\def\herwig{{\sc Herwig}}
\def\powheg{{\sc Powheg}}
\begin{document}

\title{\hspace*{1.8cm}  Parton shower contributions to  jets 
 \\ 
\hspace*{2.3cm}   from    high rapidities   at the LHC}   

\author{\hspace*{3.2cm}{\slshape M.~Deak$^{1}$,  F.~Hautmann$^2$, H.~Jung$^{3,4}$ and 
K.~Kutak$^5$}\\[1ex]
\hspace*{1.2cm}$^1$Universidade de Santiago de Compostela,   
E-15782 Santiago de Compostela\\
\hspace*{2.5cm}$^2$Theoretical Physics, 
University of Oxford,    Oxford OX1 3NP \\
\hspace*{3.0cm}$^3$Deutsches Elektronen Synchrotron, D-22603 Hamburg\\
\hspace*{3.4cm}$^4$CERN, Physics Department, CH-1211 Geneva 23\\
\hspace*{1.6cm}$^5$Instytut Fizyki Jadrowej im H. Niewodniczanskiego,  
 PL 31-342 Krakow
 }

\contribID{xy}

\doi  

\maketitle

\begin{abstract}
We  discuss current  issues associated   
with  the dependence of  jet distributions   at the 
LHC  on the behavior of 
  QCD parton  showers  for   high rapidities.\\    
\vskip 0.2 cm 
\hskip 1.1 cm 
{\em Contributed at the Workshop DIS2012, University of Bonn, March 2012}   
\end{abstract}

\vskip 0.8 cm

 At the LHC,  
  due  to 
  the phase space opening up  at  high center-of-mass energy, 
hadronic 
 jets  are   accessed for the first time  
in a    region  sensitive to contributions of high 
rapidities~\cite{ajaltouni},   in which  the forward kinematics forces 
the hard   process  into a   regime characterized by 
multiple hard scales~\cite{jhep09}.  
In this  multi-scale  region  the  production    cross section is affected by 
 high-energy  logarithmically-enhanced   corrections    
 to all orders  in the strong coupling, 
 requiring resummation methods~\cite{hef} to go beyond finite-order 
   perturbation theory.   Moreover, with increasing center-of-mass energies 
   and rapidities  
    the nonperturbative  parton  distributions  are probed  
for   smaller  longitudinal momentum fractions.   
This implies  that   effects       on   jet   distributions   
    from  multiple parton collisions~\cite{pavtrel82}  
 become more pronounced~\cite{bartal}   
     due to  the increase in  the parton density. 
     
Measurements of inclusive jet production 
are being  carried out 
at the LHC~\cite{atlas-1112,cms-11-004}  
 over  a     kinematic  range in transverse momentum 
and rapidity  much larger  than at the Tevatron and previous colliders. 
Comparisons  
 with standard model theoretical predictions   are   based 
 either  on 
next-to-leading-order (NLO) QCD calculations, supplemented with 
nonperturbative  (NP) corrections~\cite{atlas-1112,cms-11-004} estimated from 
Monte Carlo  event generators,  or on NLO-matched parton shower event 
generators of the kind  described in~\cite{ma}.

This article considers     effects of QCD  parton showers  on   jet production 
 for increasing  rapidity.  As  discussed in~\cite{jhep09,hj_rec} 
such  multi-scale   processes 
are      sensitive 
to effects of  the finite transverse-momentum 
tail of QCD multi-parton   matrix elements. 
The theoretical framework  to take these effects into account is based on 
using partonic matrix element and 
initial-state  distributions  unintegrated in 
both longitudinal and transverse momenta~\cite{mw92,skewang,hefplus}. 
On the other hand,   in NLO    event generators
   finite-k$_\perp$  terms 
 are taken into account  only partially,  
  through  the higher-order correction at  fixed  $\alpha_s$ order.    Parton
   shower  generators  
based on collinear evolution, which    are  either  matched to 
NLO calculations~\cite{ma} or   used to extract  the NP  
corrections~\cite{atlas-1112,cms-11-004}, 
  do not include 
finite-k$_\perp$ terms, as     these  terms  
  correspond to  modifications   to 
angular or transverse-momentum ordering~\cite{mw92,skewang,hefplus}.  
In what  follows we illustrate parton showering effects using 
three Monte Carlo event generators: 
 the  k$_\perp$-shower \cascade\  generator~\cite{cascadedocu},  
  the NLO matched \powheg\  generator~\cite{alioli}, 
and 
   \pythia\  shower  Monte Carlo~\cite{pz_perugia},  
 used in two different modes:  with    the  tune P1~\cite{pz_perugia}  
 including    multiple parton  collisions, 
and with single parton collision (\pythia-nompi).  
As emphasized  in~\cite{preprint},  
effects   coming from  
noncollinear  multi-parton  emission    influence    jets  at large rapidities  
  as well as    jets  produced centrally but in  
association  with   observed forward final states.

\begin{figure}[htb]
\vspace{100mm}
\includegraphics{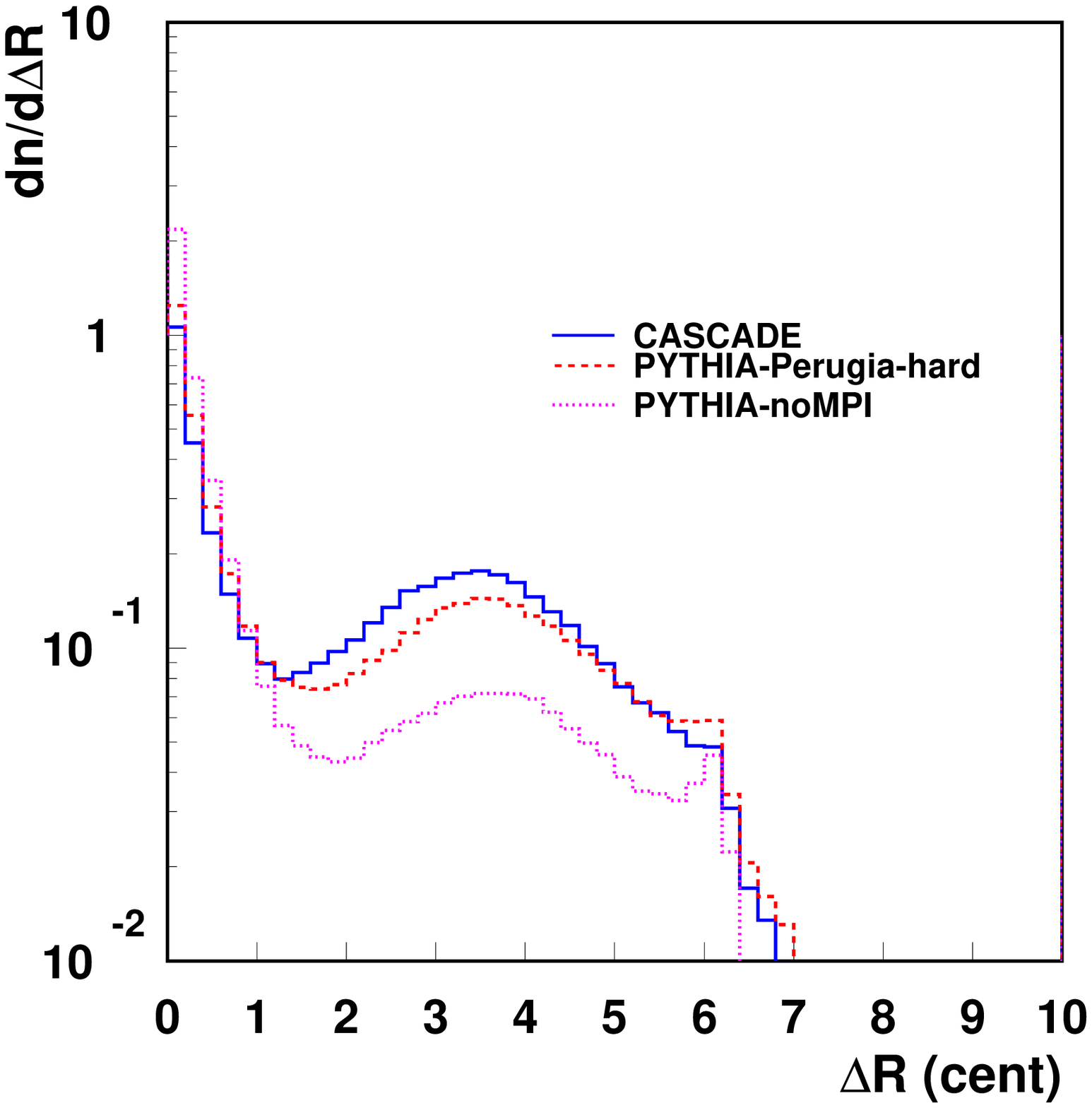}
\includegraphics{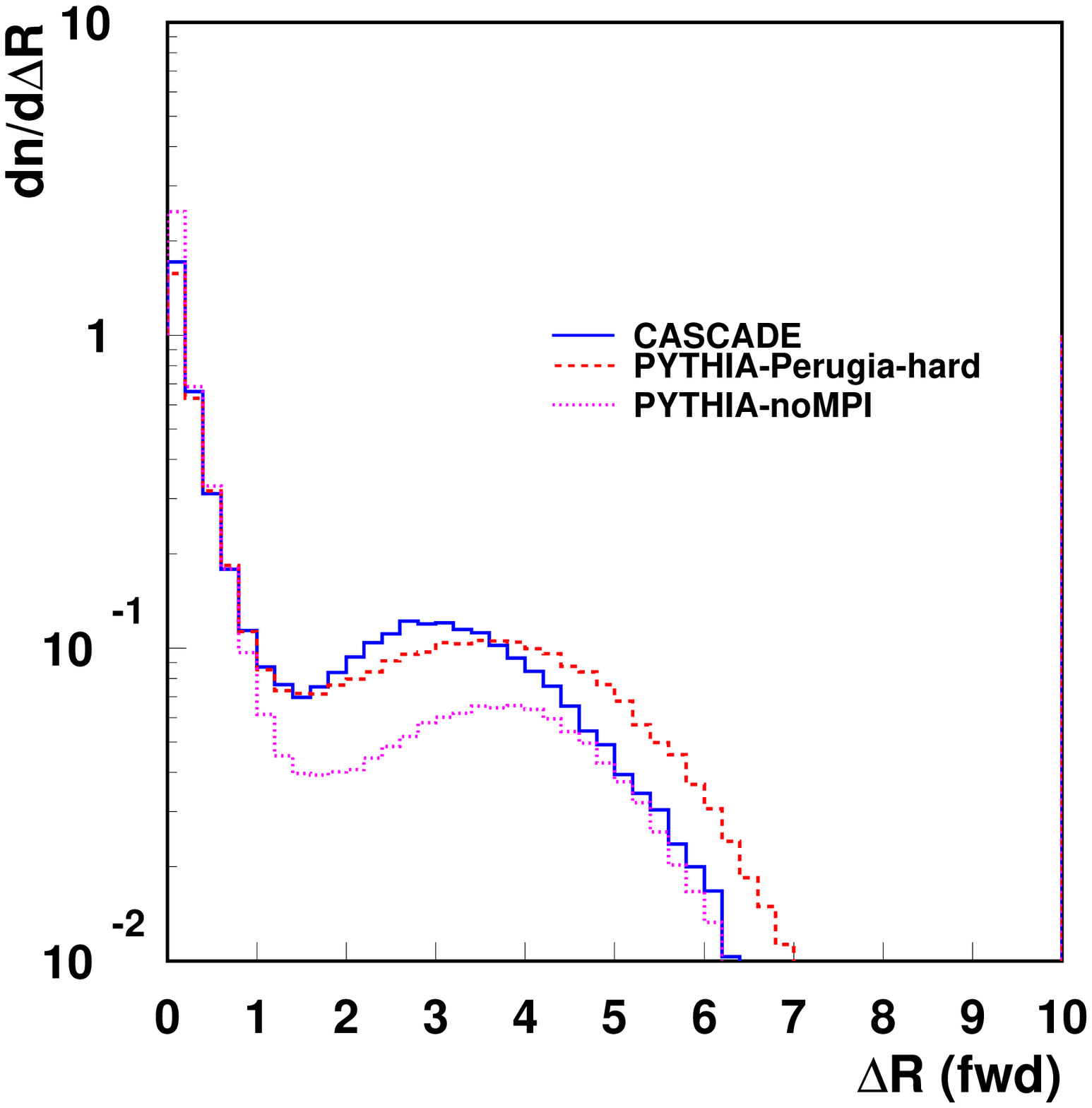}
\includegraphics{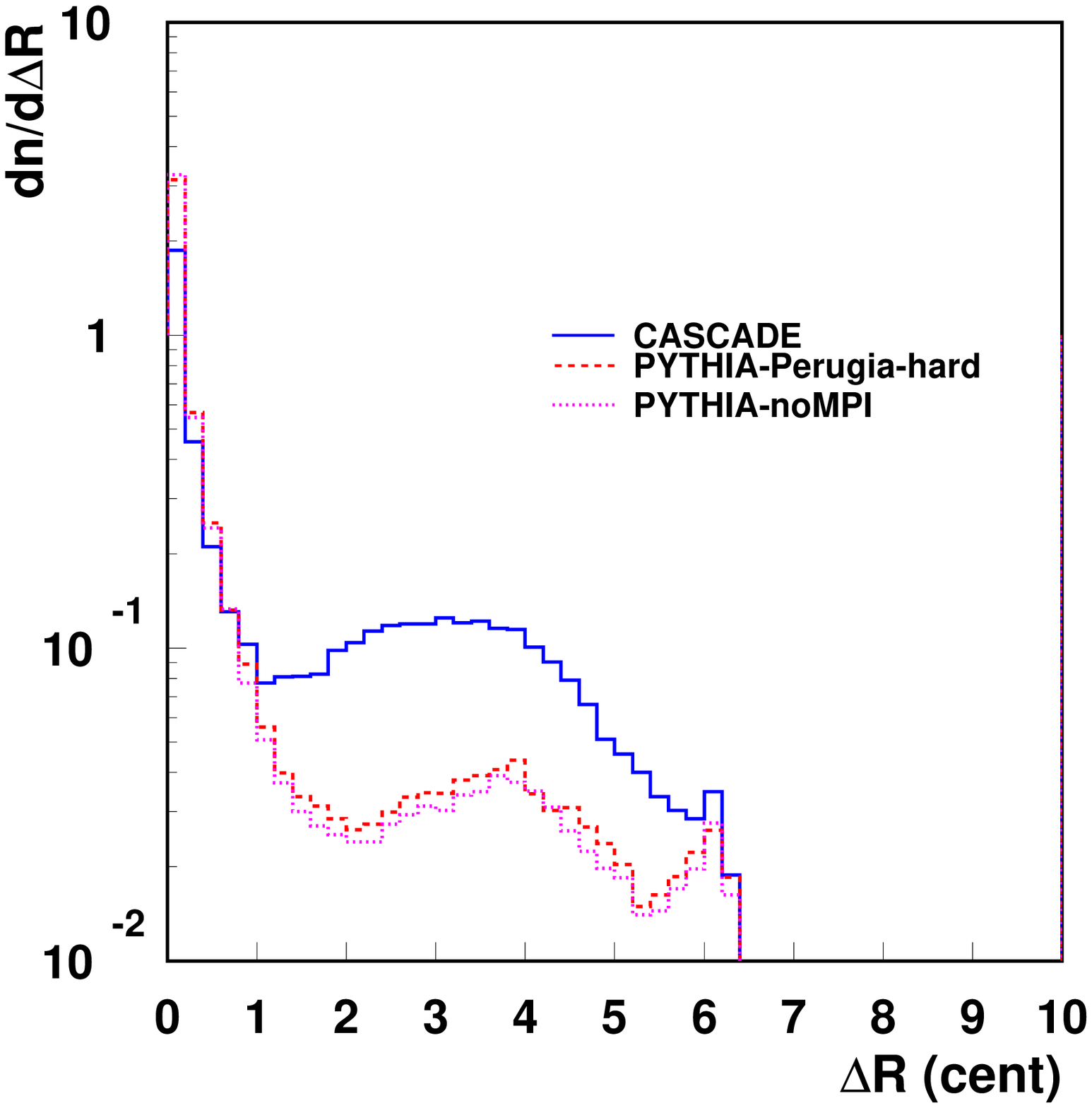}
\includegraphics{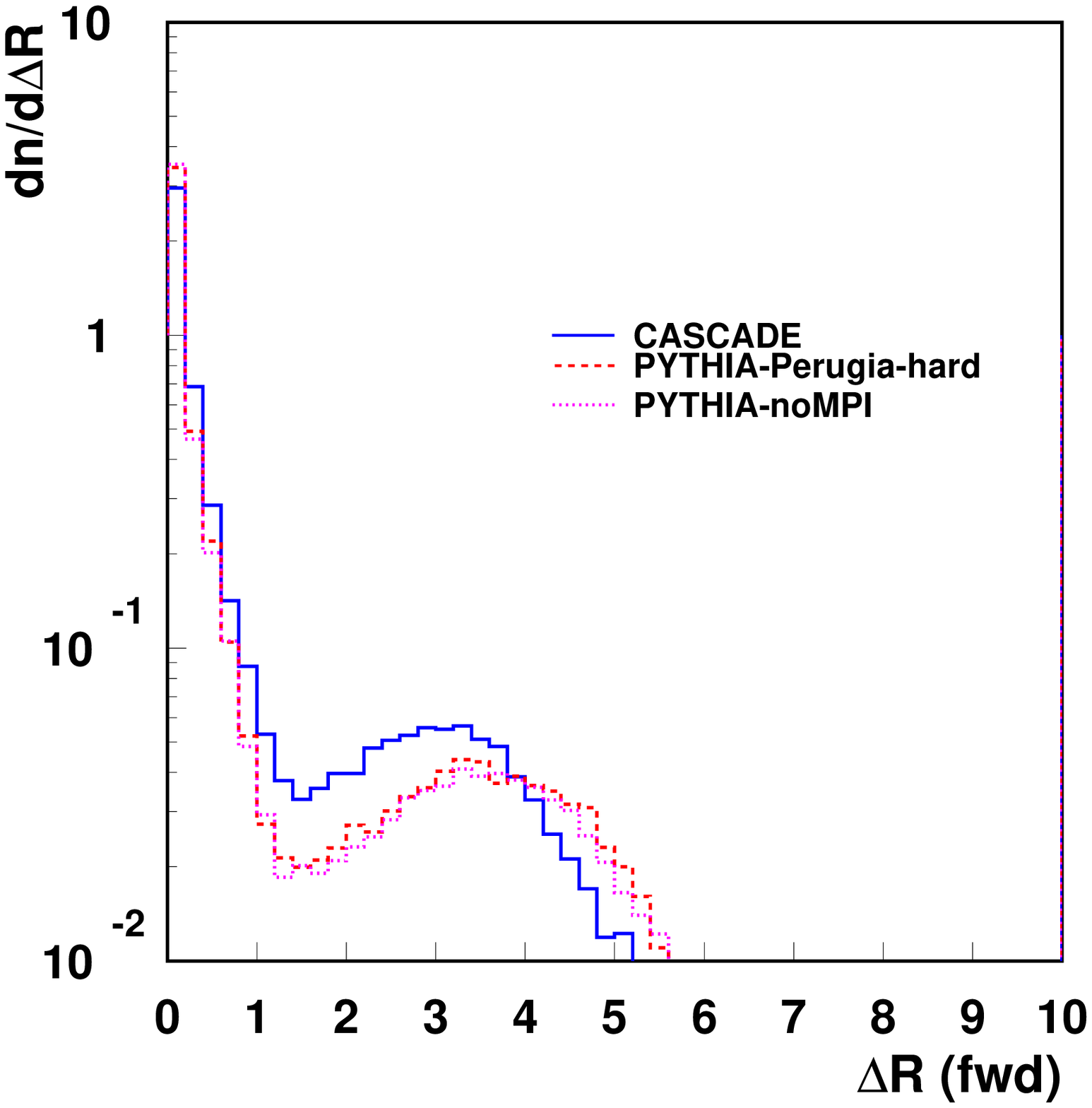}
\caption{ $\Delta R$ 
distribution of   central  ($|\eta_c|<2$, left)  and forward jets ($3 < |\eta_f| <5 $, right) 
 for $E_T > 10$~GeV (upper row) and  $E_T>\!30$~GeV (lower row)~\cite{preprint}. The 
 curves 
 correspond to the 
k$_\perp$-shower  Monte Carlo generator     \cascade\   and to    
the  \pythia\  shower Monte Carlo generator  used in two modes, 
one  in which  multiple parton interactions  are included and  one  in which they are 
switched off.} 
\label{fig:deltar}
\end{figure}

In    Fig.~\ref{fig:deltar} we consider final states associated  
with production of a  forward and a central jet~\cite{preprint} 
reconstructed via  the Siscone algorithm~\cite{siscone} ($R= 0.4$) 
and report the 
$\Delta R =   \sqrt{\Delta \phi^2 + \Delta \eta^2 }$ distribution,  
 where 
 $\Delta \phi= \phi_{jet} - \phi_{part}$ ($\Delta \eta= \eta_{jet} - \eta_{part}$) is the azimuthal 
 (rapidity) difference between the jet and the 
 corresponding parton from the matrix element.      This distribution probes to what extent 
 jets are dominated by hard partons in the matrix element  or  
 originate   from the  showering. The  large-$\Delta R$  region,  
 corresponding to sizeable contributions to jets from showers, 
   is  seen     
 to be enhanced by noncollinear corrections.  While this  effect  can be also  produced 
   by 
 multi-parton interactions    for   low $E_T$ jets, this  no longer applies  as $E_T$  
  increases.   
   It is noteworthy that as a 
   consequence  of  high-rapidity  correlations   
  the  enhanced  dependence of jet distributions on features of the  parton showers   
  is especially pronounced for central jets.

\begin{figure}[htbp]
\vspace{70mm}
\includegraphics{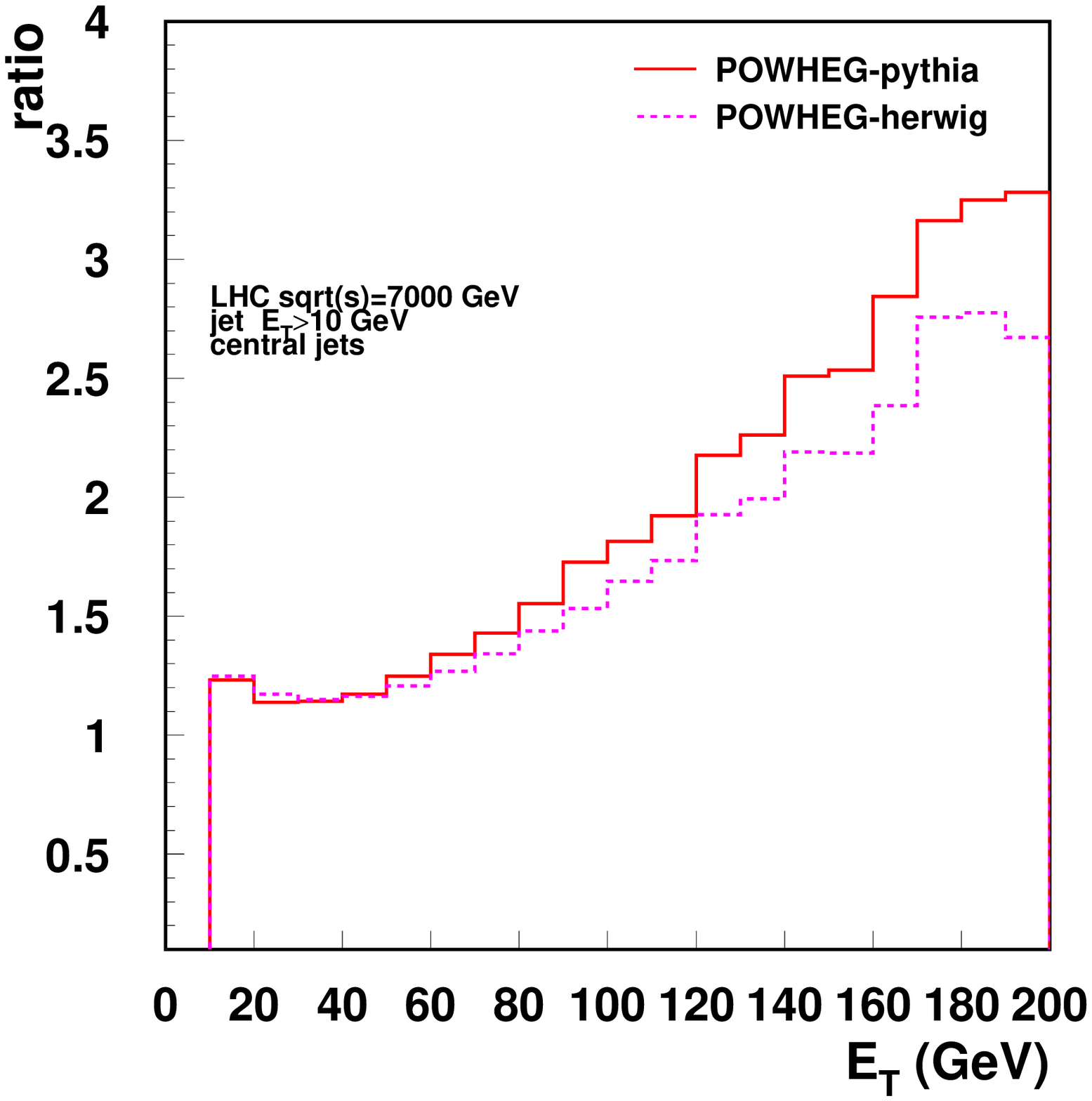}
\includegraphics{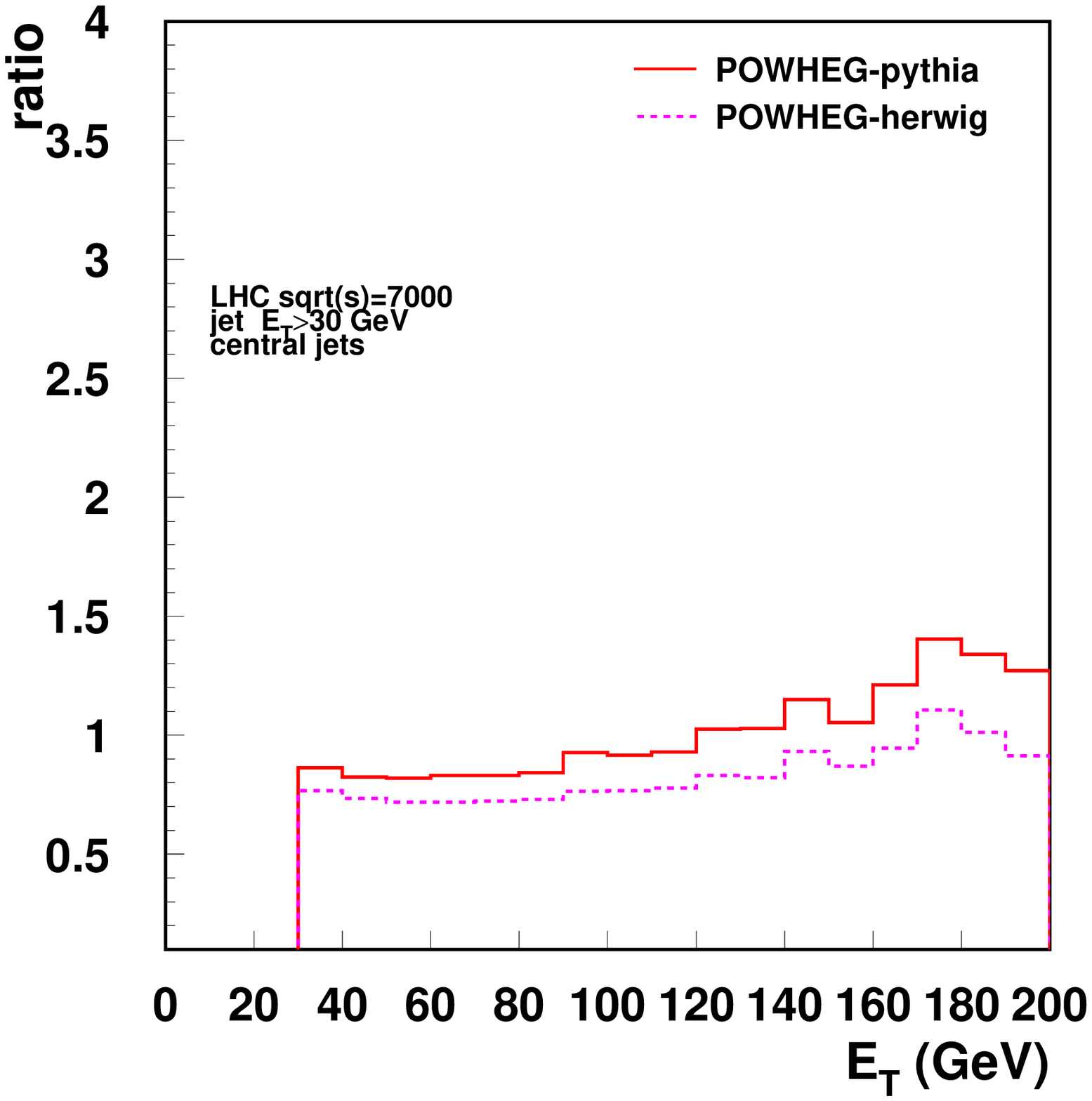}
\caption{Ratio of  central jet transverse energy 
spectra   from the NLO-matched  Monte Carlo generator  \powheg,   
 interfaced with   \pythia\ and \herwig\  parton  showers, to the no-showering result. 
(left) $E_T > 10$~GeV; (right) $E_T>\!30$~GeV. } 
\label{fig:powhegshowers}
\end{figure}

In Fig.~\ref{fig:powhegshowers}   this issue  is examined using the NLO event 
generator \powheg\  matched with   parton showers      \pythia\ and \herwig. 
We show the central jet transverse energy spectrum for the two cases, 
normalized to the result obtained by  switching off parton showering.  The 
marked differences between the two cases  are consistent 
with the findings in~\cite{cms1202}, and with the  large  
contribution to  jets from  showering  indicated by    Fig.~\ref{fig:deltar}.  
In particular   this  suggests  
 that high-rapidity  correlations   affect  the behavior  of 
jet distributions in the central region.

We observe  that  while first  measurements of forward 
jet spectra~\cite{cms1202} are roughly in agreement 
with   Monte Carlo   simulations,   detailed 
aspects of production rates   and  correlations~\cite{cms1202,dijetratios}    
are not  well    understood yet.  Also,    
hadronic event shapes measured at the LHC~\cite{evshape-cms}    
suggest    that   parton showering  effects  dominate 
   contributions of hard matrix elements   evaluated at high multiplicity. 
The numerical results~\cite{preprint}  for the  
large rapidity region 
underline 
especially the significance of  contributions  to  showering   from transverse 
momentum  dependent  
branching~\cite{unint09} and parton  distributions~\cite{jccfh01}. 
This  region is  relevant  to many  aspects of LHC  physics, including   
 studies of jets from   
decays of  highly boosted new  particles~\cite{jetsubs},  new particle 
 searches  using 
vector boson fusion channels~\cite{vvfusion}, 
 relationship  of forward particle production and cosmic ray physics~\cite{grotheetal}, 
high-density QCD and heavy ion collisions~\cite{dent}. 
The treatment in terms of  
unintegrated distributions 
may in particular be useful 
to  investigate effects of gluon rescattering~\cite{rescatt}  within 
parton branching approaches.

\vskip 0.5 cm 

\noindent 
{\bf Acknowledgments}.   We  thank  the conveners    for the invitation  and 
excellent organization of the  meeting.


{\raggedright
\begin{footnotesize}



\end{footnotesize}
}


\end{document}